\documentclass[a4paper/letterpaper,
               biblatex,     % biblatex is used
               keeplastbox,   % flushend option: not to un-indent last line in References
               ]{jacow}
\usepackage{pdfpages,multirow,ragged2e} %

\usepackage{smartdiagram}

\usepackage{tikz}
\usetikzlibrary{shapes.geometric, arrows,backgrounds,shapes.arrows}

\tikzstyle{invisible} = [rectangle, minimum width=0.1cm, minimum height=1cm, text centered, draw=none ]
\tikzstyle{hardware2} = [rectangle, minimum width=0.5cm, minimum height=0.5cm, text centered, text width=1cm,draw=black, fill=blue!30]

\tikzstyle{ioc} = [rectangle, minimum width=1cm, minimum height=1cm, text centered, draw=black, fill=orange!60,rounded corners]

\tikzstyle{tca} = [rectangle ,minimum width=1cm, minimum height=1cm, text centered, draw=black, fill=red!30]
\tikzstyle{serviceg} = [rectangle, minimum width=1cm, minimum height=1cm, text centered, draw=black, fill=green!30, rounded corners]
\tikzstyle{serviceaq} = [rectangle, minimum width=1cm, minimum height=1cm, text centered, draw=black, fill=teal!40, rounded corners]
\tikzstyle{serviceda} = [rectangle, minimum width=1cm, minimum height=1cm, text centered, draw=black, fill=yellow!30, rounded corners]

\tikzstyle{gui2} = [rectangle, minimum width=1cm, rounded corners,minimum height=1cm, text centered, draw=black, fill=violet!30]

\tikzstyle{arrow2} = [thick,<->,>=stealth]

\tikzstyle{hardware} = [rectangle, minimum width=3cm, minimum height=1cm, text centered, draw=black, fill=blue!30]
\tikzstyle{linux} = [rectangle ,minimum width=3cm, minimum height=1cm, text centered, draw=black, fill=red!30]
\tikzstyle{service} = [rectangle, minimum width=3cm, minimum height=1cm, text centered, draw=black, fill=green!30]
\tikzstyle{gui} = [rectangle, minimum width=3cm, rounded corners,minimum height=1cm, text centered, draw=black, fill=orange!30]

\tikzstyle{decision} = [diamond, minimum width=3cm, minimum height=1cm, text centered, draw=black, fill=green!30]

\tikzstyle{arrow} = [thick,->,>=stealth]

\makeatletter%
	\ifboolexpr{bool{xetex}}
	 {\renewcommand{\Gin@extensions}{.pdf,%
	                    .png,.jpg,.bmp,.pict,.tif,.psd,.mac,.sga,.tga,.gif,%
	                    .eps,.ps,%
	                    }}{}
\makeatother

\ifboolexpr{bool{xetex} or bool{luatex}} % test for XeTeX/LuaTeX
 {}                                      % input encoding is utf8 by default
 {\usepackage[utf8]{inputenc}}           % switch to utf8

\usepackage[english]{babel}

\begin{document}

\title{Software Development for APS-Upgrade LLRF}

%%\author{A. N. Author\thanks{email address}, H. Coauthor, Name of Institute or Affiliation, City, Country \\
%%		P. Contributor\textsuperscript{1}, Name of Institute or Affiliation, City, Country \\
%%		\textsuperscript{1}also at Name of Secondary Institute or Affiliation, City, Country}

%%\author[1]{T. Madden\thanks{tmadden@anl.gov}}
%%\author[1]{N. Arnold} 
%%\author[1]{T. Berenc} 
%%\author[2]{E. Breeding}  
%%\author[1]{T. Fors} 
%%\author[1]{G. Shen}  
%%\author[1]{S Veseli} 
%%\author[1]{Y. Yang} 
%%\author[2]{K.  Vodopivec}
%%\author[1]{W. Yoder }
%%\affil[1]{Argonne National Laboratory, Chicago,IL, USA}
%%\affil[2]{Oak Ridge National Laboratory, Oak Ridge, TN, USA}

\author{T. Madden\thanks{tmadden@anl.gov}, N. Arnold, T. Berenc,   T. Fors, G. Shen,  S. Veseli, Y. Yang,  W. Yoder\\
Argonne National Laboratory, Chicago, IL, USA\thanks{Argonne National Laboratory's work was supported by the U.S. Department of Energy, Office of Science, under contract DE-AC02-06CH11357.} \\
\\
 E. Breeding, K. Vodopivec\\
Oak Ridge National Laboratory, Oak Ridge, TN, USA}

\maketitle

\begin{abstract}

An extensive service-based software system has been developed for real-time data acquisition (DAQ) on numerous subsystems of the Advanced Photon Source (APS) accelerator.
  The software, called DAQ, is based upon EPICS 3 Area- Detector with added 
EPICS 4 services, with EPICS 3 + 4 being called EPICS 7\cite{rivers2010areadetector},\cite{epics4} . DAQ is 
comprised of several network services including file storage and 
processing as well as numerous EPICS 7 Input-Output Controller (IOC) applications 
that collect data and publish to EPICS 7 from accelerator subsystems, including 
power supplies, diagnostics, timing, and RF systems\cite{veseli2019}. For the APS upgrade 
extensions to the DAQ software have been developed for digital Low Level RF 
(LLRF) systems supporting multiple hardware platforms including Lawrence Berkeley Laboratory's 
LLRF4 hardware and Micro Telecommunications Computing Architecture ($\mu$TCA) \cite{vadatech}.  EPICS 7 IOCs for LLRF run on multiple 
threads for real-time data collection, visualization , system control, 
and data file storage that can be triggered on accelerator beam dumps for 
system debug. A Python-based oscilloscope application has been developed for 
real-time data visualization.

\end{abstract}

\section{Introduction}

The APS at Argonne is being upgraded from its current configuration to a new Multi-Bend Achromat (MBA) lattice, called APSU \cite{apsu}. The LLRF systems are being 
upgraded from analog circuitry to new digital LLRF. The new digital LLRF
will be comprised of $\mu$TCA hardware, with a crate housing several 
modules, including an FPGA-based Advanced Mezzanine Card (AMC) from Vadatech,
Inc.  with custom LLRF daughter boards, and a Linux blade to run control and
data acquisition software \cite{vadatech}. While the main engineering efforts
for the LLRF upgrade are in signal processing, firmware and software design, 
we present here a summary of the software development.

The LLRF software is one part of a larger software data acquisition and 
control infrastructure, collectively called DAQ.
A brief overview of DAQ and its LLRF-specific components will be presented.
The APSU LLRF system will support three separate machines in the APSU: Particle
Accumulation Ring, Synchrotron Booster, and Storage Ring. A common software, hardware and firmware
platform will be used for all three machines.

\section{Software Infrastructure}

\subsection{EPICS 7}
The digital control system used at APS is Experimental Physics and 
Industrial Control System (EPICS). EPICS is a software tool set that allows 
the digital control of large systems such as a particle accelerator. The EPICS
system at APS includes a large collection of Input-Output Controllers (IOC), or 
networked computer servers  that interface to hardware and publish hardware 
settings and controls as Process Variables (PVs). 
EPICS clients such as GUI applications allow the monitoring and control of the
hardware components by network connection to multiple PVs. 
While the current version of EPICS for
process control is EPICS version 3, a major addition is EPICS version 4
which allows the streaming of structured data over a network. The
combination of EPICS 3 and 4 is called EPICS 7. 
 
\subsection{Area Detector + EPICS 7}

Area Detector (AD) is a software tool set for EPICS control of large area
detectors at synchrotron beam lines \cite{rivers2010areadetector}. In
particular, AD provides a system for acquiring data from large detectors using
a "driver" that interfaces to hardware, collects data,  and passes these data
to a set of queues. Other components in AD, called "plug-ins," retrieve these
images from queues and perform real-time processing before saving image data to disk.
The DAQ software IOC is an extension of the AD IOC, in that EPICS 4 data
structures relevant to accelerators, rather then images,  are passed through
the system of queues. AD's set of queues, multi-threading, and ability to pass
large datasets in real time makes it attractive for use as a DAQ for
accelerator systems.  

\subsection{Data Acquisition System (DAQ)}

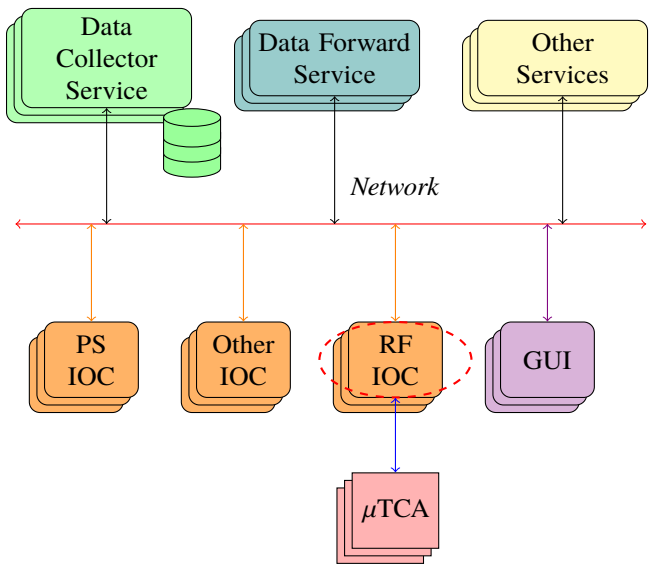
\begin{figure}
\centering

\begin{tikzpicture}[node distance=1.5cm]

\node foreach \ofst in {0.0,0.1,0.2} (fsave) at ([xshift=\ofst cm,yshift=\ofst cm]0cm,0cm) [serviceg,text width=2cm] {Data Collector Service};

\node foreach \ofst in {-0.40,-0.20,0.0} (disk) at ([yshift=\ofst cm]fsave.south east) [cylinder, shape border rotate = 90, draw, fill=green!40, minimum width = .75cm,minimum height = 0.3cm,anchor=north] {};

\node foreach \ofst in {0.0,0.1,0.2} (dfwd) at ([xshift=\ofst cm,yshift=\ofst cm]3cm,0cm) [serviceaq,text width=2cm] {Data Forward Service};

\node foreach \ofst in {0.0,0.1,0.2} (oths) at ([xshift=\ofst cm,yshift=\ofst cm]6cm,0cm) [serviceda,text width=2cm] {Other Services};

\node at (4cm,-1.5cm) [] {\textit{Network}};

\draw[<->,red] (-1.0cm,-2cm) -- (7.3cm,-2cm);

\draw[<->,black] (fsave) -- ++(0cm,-2.2cm);
\draw[<->,black] (dfwd) -- ++(0cm,-2.2cm);
\draw[<->,black] (oths) -- ++(0cm,-2.2cm);

\coordinate (c0) at (-0.2cm,-4cm);
\coordinate (c1) at ([xshift=2cm]c0); 
\coordinate (c2) at ([xshift=2cm]c1); 
\coordinate (c3) at ([xshift=2cm]c2); 
\coordinate (c4) at ([yshift=-2cm]c2); 

\node foreach \ofst in {0.0,0.1,0.2} (psioc) at ([xshift=\ofst cm,yshift=\ofst cm]c0) [ioc,text width=1cm] {PS IOC};

\node foreach \ofst in {0.0,0.1,0.2} (othioc) at ([xshift=\ofst cm,yshift=\ofst cm]c1) [ioc,text width=1cm] {Other IOC};

\node foreach \ofst in {0.0,0.1,0.2} (rfioc) at ([xshift=\ofst cm,yshift=\ofst cm]c2) [ioc,text width=1cm] {RF IOC};

\node foreach \ofst in {0.0,0.1,0.2} (guix) at ([xshift=\ofst cm,yshift=\ofst cm]c3) [gui2,text width=1cm] {GUI};

\draw[red,thick,dashed] ([xshift=.2cm,yshift=.2cm]c2) circle (1cm and 0.5cm);
\draw[<->,orange] (psioc) -- ++(0cm,1.8cm);
\draw[<->,orange] (othioc) -- ++(0cm,1.8cm);
\draw[<->,orange] (rfioc) -- ++(0cm,1.8cm);
\draw[<->,violet] (guix) -- ++(0cm,1.8cm);

\node foreach \ofst in {0.0,0.1,0.2} (utca) at ([xshift=\ofst cm,yshift=\ofst cm]c4) [tca] {$\mu$TCA};

\draw[<->,blue] (utca) -- (rfioc);
\end{tikzpicture}

\caption{DAQ System showing Network Services, IOCs, and GUI Client. The LLRF IOC is circled.}
\label{fig:daqsys}
\end{figure}

The DAQ system is a large distributed software system at the APS for data
acquisition of accelerator data. DAQ is comprised of a network of EPICS 7 IOCs that
interface to accelerator system hardware and collect and stream data over the
network. DAQ also includes a set of network services for data storage, data
routing, data processing and several other applications. Additionally, DAQ
includes an interface to Self Describing Data Set (SDDS), a structured data file type used at APS for
storing accelerator data\cite{sdds}. Finally DAQ can interface to several mathematical
software tools such as Python and
Octave\cite{octave}. A simplified diagram of DAQ can be
seen in Figure \ref{fig:daqsys}.

%\begin{figure}
%\includegraphics[width=\linewidth]{daqdia2.png}
%\caption{DAQ System. Many IOCs and network services comprise DAQ. The LLRF IOC interfaces to RF hardware.}
%\label{fig:daqsys}
%\end{figure}

\section{EPICS 7 LLRF IOC}

The new digital LLRF system for the APSU utilizes an AD-based  DAQ IOC that
not only controls and monitors the LLRF system, but also collects and streams
data to various servers and clients on the network in real time.  For the LLRF
IOC, the driver gathers RF data from the APSU RF systems and passes them raw to a
set of queues. DAQ plug-ins then take the raw data off the queue, format them as
EPICS 4 data structures,  perform real-time processing on the data before
finally streaming the data to a networked file service for storage.  The
drivers and plug-ins can run on multiple threads in one process on a Linux
computer. The networked data storage service saves RF data on one centralized
server rather than at local storage scattered around the APS site. Centralized
data storage allows for long term secure storage as well as centralized data
processing.

\subsection{$\mu$TCA PCIe Interface}

The LLRF $\mu$TCA crate houses a Linux blade running the EPICS 7 LLRF IOC as
well as an FPGA-based RF card that generates and captures the RF signals in the
LLRF system. Communication between the AMC and Linux blade occurs via a PCIe x4
interface.  The FPGA runs firmware that implements a PCIe device on the bus.
This firmware, developed by Oak Ridge National Laboratory(ORNL), allows for register-based read and writes to control "knobs" in
the digital LLRF system, as well has Direct Memory Access (DMA) PCIe transfers
for streaming real-time RF data from FPGA to the Linux blade. 

\begin{figure}
\centering
\begin{tikzpicture}[node distance=0.75cm]

\node (tcahardware) [shape=rectangle,draw,minimum height = 0.75cm,minimum width = 4cm,fill=blue!30] {$\mu$TCA Hardware};
\node (invis) [shape=rectangle,minimum height = 0.75cm,minimum width = 4cm,above of = tcahardware] {};
\node (occmod) [shape=rectangle,draw,minimum height = 0.75cm,minimum width = 4cm,fill=red!30,above of = invis] {OCC Kernel Module};
\node (kernel) [shape=rectangle,draw, above of= occmod,minimum height = 0.75cm,minimum width = 4cm,fill=red!30] {Linux Kernel};
\node (occlib) [shape=rectangle,draw,above of = kernel,minimum height = 0.75cm,minimum width = 4cm,fill=red!30] {OCC App Library};
\node (driver) [shape=rectangle,draw,above of= occlib,minimum height = 0.75cm,minimum width = 4cm,fill=red!50] {DAQ Driver};
\node (plg1) [shape=rectangle,draw,above of=driver,minimum height = 0.75cm,minimum width = 4cm,fill=red!50 ] {DAQ Proc. Plug-in};
\node (plg2) [shape=rectangle,draw,above of=plg1,minimum height = 0.75cm,minimum width = 4cm,fill=red!50 ] {DAQ Stream Plug-in};

\draw [<->] (tcahardware) -- (occmod);
%\draw [arrow] (ss) -- (pycode);
%\draw [arrow] (pycode) -- (make);
%\draw [arrow] (make) -- (done);

\end{tikzpicture}

\caption{Software Stack in Linux blade shown in red. The Linux Kernel uses 
the OCC module to communicate to the $\mu$TCA hardware via PCIe. The EPICS IOC
Application, shown in darker red,  calls the OCC library and is comprised of a Driver
and a set of Plug-ins. }
\label{fig:swstack}
\end{figure}
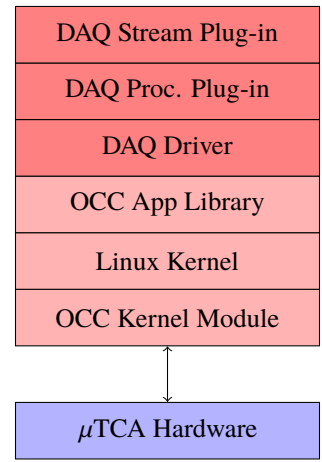

The Linux blade runs a custom Linux Kernel Module, called OCC and developed by ORNL,
that handles all transactions on the PCIe bus from the FPGA
card\cite{vodo}. The EPICS 7 LLRF IOC runs in the application layer and makes calls to the
OCC application library to communicate with the FPGA card. For debugging, the
FPGA card can be controlled at the Linux command line by using the OCC command
line tools. In this way, system problems can be investigated by shutting off
the EPICS IOC, and directly controlling the hardware over PCIe.  A diagram of
the software stack is shown in Figure \ref{fig:swstack}.

\subsection{Automatic Screen and Code Generation}

To control and read back the LLRF system, the IOC application reads and writes
to FPGA registers addressed as memory via the PCIe bus. Because the FPGA
firmware is under constant development independent of the software, a method
was devised for keeping the software synchronised with FPGA firmware. All of
the FPGA registers are documented in a spreadsheet that includes register
addresses, bit fields in the registers, and whether or not the registers need
to be EPICS PVs.  For example, at register address 0xC00, bit 0 could be
defined as turning on and off RF output from the $\mu$TCA crate.  Whenever the
firmware is updated, the spreadsheet is updated to document firmware changes. 
 
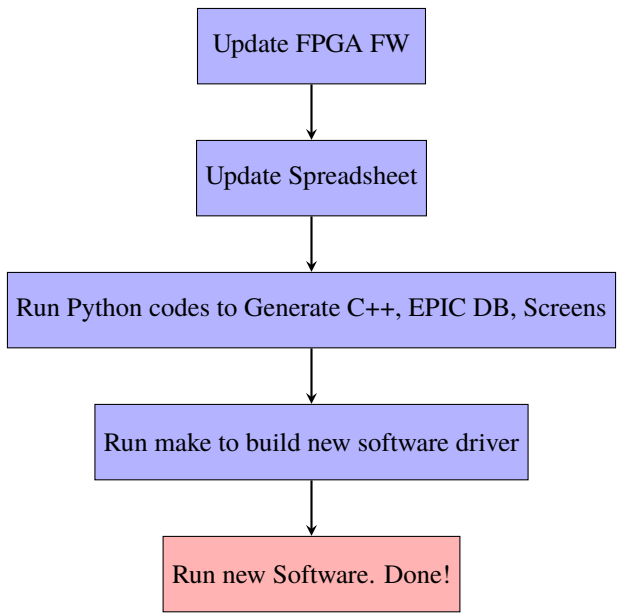
\begin{figure}
\centering
\begin{tikzpicture}[node distance=1.75cm]

\node (upfpga) [hardware] {Update FPGA FW};
\node (ss) [hardware, below of=upfpga] {Update Spreadsheet};
\node (pycode) [hardware, below of=ss] {Run Python codes to Generate C++, EPIC DB, Screens};
\node (make) [hardware, below of=pycode] {Run make to build new software driver};
\node (done) [linux, below of=make] {Run new Software. Done!};

\draw [arrow] (upfpga) -- (ss);
\draw [arrow] (ss) -- (pycode);
\draw [arrow] (pycode) -- (make);
\draw [arrow] (make) -- (done);

\end{tikzpicture}

\caption{Work flow for upgrading FPGA and Software.}
\label{fig:fpga2code}
\end{figure}

A set of python scripts has been written to read the spreadsheet and
automatically generate C++ code, EPICS database files defining PVs, and GUI
screens. In this way, when a new register is added to the FPGA firmware and
documented in the spreadsheet, the python scripts can be run to automatically
update the software source code and GUI screens. Once the python scripts are
run, the IOC source code is recompiled, and the software is then able to
control the new registers. By using the spreadsheet and python scripts,
software and screens need not be updated by hand every time a new bit is added
to the FPGA. The tool flow is shown in Figure \ref{fig:fpga2code}.

%\begin{figure}
%\smartdiagramset{font=\footnotesize,module minimum width=2.2cm,module x sep=2.5,text width=2.0,back arrow disabled=true}
%\smartdiagram[flow diagram:horizontal]{Update FPGA FW, Update Spreadsheet, Run Py Codes, Compile, Run New GUI and IOC}
%\smartdiagram[priority descriptive diagram]{FPGA,PCIe Transfer,DAQ Driver grab packets,DAQ Plugin parse to EPICS4 ,Stream EPICS4 to network,File Store and/or Real-time Display}

%\end{figure}

\subsection{Streaming Signals from the LLRF Control Loop}

\begin{figure}
\centering
\begin{tikzpicture}[node distance=1.75 cm]

\node (cavadc) [shape =single arrow , draw,single arrow head extend = 0,shape border rotate=180] {ADC\textsubscript{2}};
\node (refadc) [shape =single arrow , draw,single arrow head extend = 0,shape border rotate=180,above of = cavadc] {ADC\textsubscript{1}};

\node (ddc1) [shape =rectangle, draw,right of = cavadc] {DDC};
\node (ddc2) [shape =rectangle, draw,above of = ddc1] {DDC};

\node (adder) [shape =circle, draw,right of = ddc1] {+};
\node (minus) [right of = ddc1,xshift=-0.4cm, yshift=-0.3cm] {-};
\node (setpt) [shape =rectangle, draw,above of = adder] {SetPoint};
\node (pid) [shape =rectangle, draw,right of = adder] {PID};

\node (duc) [shape =rectangle, draw,below of = ddc1] {DUC};

\node (dac) [shape =single arrow , draw,single arrow head extend = 0,right of = duc] {DAC};

%\node (fbout) [above of =ddc1, xshift=1cm,yshift=-1cm] {FB I/Q};

\draw [arrow] (cavadc) -- node [anchor=south,yshift=0.1cm]{\textcolor{red}{\textit{Raw}}}  (ddc1);
\draw [arrow] (refadc) -- node [anchor=south,yshift=0.1cm]{\textcolor{red}{\textit{Raw}}}  (ddc2);
\draw [arrow] (ddc1) -- node [anchor=south,yshift=0.1cm]{\textcolor{red}{\textit{FB I/Q}}} (adder);
\draw [arrow] (ddc2) -- node [anchor=south,yshift=0.1cm]{\textcolor{red}{\textit{Ref I/Q}}} (setpt);
\draw [arrow] (setpt) -- node [anchor=west,yshift=0.1cm]{\textcolor{red}{\textit{SP I/Q}}}  (adder);
\draw [arrow] (adder) --  node [anchor=south,yshift=0.1cm]{\textcolor{red}{\textit{Err I/Q}}} (pid);
\draw [arrow] (pid) --  node [anchor=north west]{\textcolor{red}{\textit{Drv I/Q}}} (duc);
\draw [arrow] (duc) -- (dac);

\end{tikzpicture}

\caption{LLRF Control Loop showing streamed signals in red. }
\label{fig:pidloop}
\end{figure}
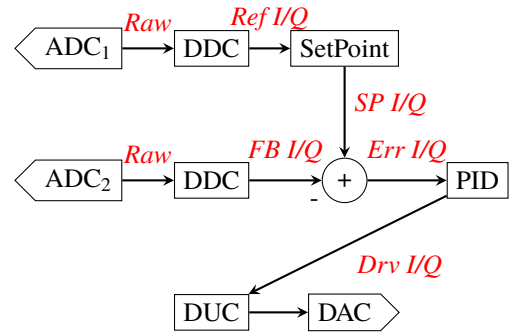

The LLRF system is a Proportional-Integral-Derivative (PID) loop  to control
the magnitude and phase of the RF signal in accelerator cavities, as measured
by cavity probes. While the current analog APS LLRF system directly controls
magnitude and phase, the new digital LLRF system splits the RF signal into I
and Q components, and runs a separate PID controller for each component. The
software system allows control of all settings in the PID controller. Also, the
LLRF firmware and software stream raw data from several points in the PID
control loop for real-time monitoring, data collection,and analysis of the PID
loop. A diagram of the loop, and points where data is collected by the software
system is shown in Figure \ref{fig:pidloop}.

In Figure \ref{fig:pidloop}, ADC\textsubscript{2} digitizes the cavity probe
signal which is digitally down-converted in sample rate (DDC) to create the
feedback I and Q signals (FB I/Q). Typically, the ADC samples at a sample rate of 41.5MHz, and the Digital Down Converter (DDC) converts the sample rate to 2.4MHz. A further decimated version of the signal is 
streamed through software. ADC\textsubscript{1} digitizes a reference
phase that is converted into I and Q (Ref I/Q) and is used to calculate the
set point (SP I/Q). The feedback signal is subtracted from the set point to
generate the error signal (Err I/Q) that is fed into the PID controller. The
output of the PID, the Drive signal (Drv I/Q), is digitally up-sampled and
converted to an analog RF signal to drive the cavity. The raw ADC signals can
optionally be streamed as a diagnostic.

\subsection{Software Data Stream}

The LLRF DAQ is designed to simultaneously stream I/Q data in real-time for 12 sampled waveforms.  The I/Q waveforms in the PID controller are inserted into data packets to be sent over the PCIe bus for DMA transfer into the Linux blade. The OCC kernel module handles the DMA transfer on the Linux blade, and allows the DAQ EPICS IOC driver to read the packets with a memory copy by calling the OCC library and place them into a queue.  The thread running the driver is set to a high priority and is assigned its own CPU core to minimize the chance of losing packets. Further, OCC supports "interrupt coalescing," or the grouping of many PCIe packets into one software interrupt, to reduce software overhead in the Linux blade. 

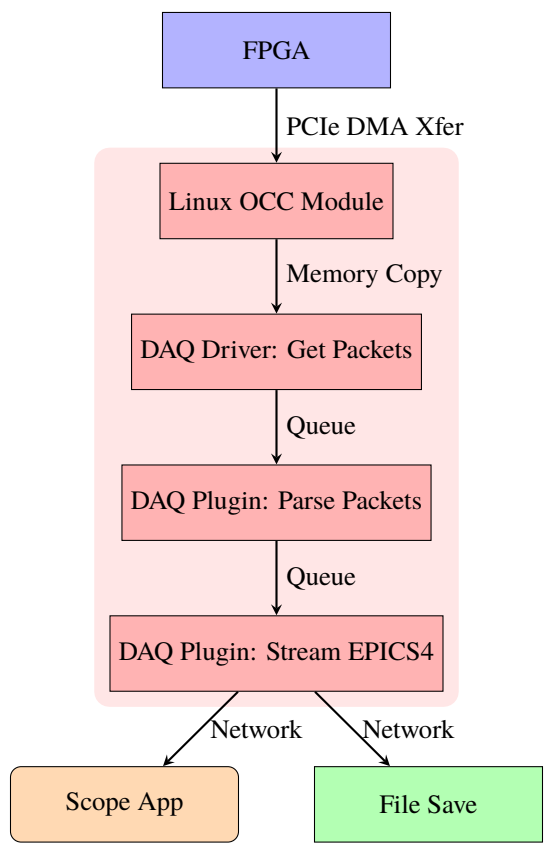
\begin{figure}
\centering
\begin{tikzpicture}[node distance=2.0cm]

\node (fpga) [hardware] {FPGA};
\node (occ) [linux, below of=fpga] {Linux OCC Module};
\node (driver) [linux, below of=occ] {DAQ Driver: Get Packets };
\node (plugin) [linux, below of=driver] {DAQ Plugin: Parse Packets};
\node (plugin2) [linux, below of=plugin ] {DAQ Plugin: Stream EPICS4};
\node (filestore) [service, below of=plugin2, xshift=2cm] {File Save};
\node (scope) [gui, below of=plugin2,xshift=-2cm] {Scope App};

\draw [arrow] (fpga) --node[anchor=west]{PCIe DMA Xfer} (occ);
\draw [arrow] (occ) --node[anchor=west]{Memory Copy} (driver);
\draw [arrow] (driver) --node[anchor=west] {Queue} (plugin);
\draw [arrow] (plugin) --node[anchor=west] {Queue} (plugin2);
\draw [arrow] (plugin2) --node[anchor=west] {Network} (filestore);
\draw [arrow] (plugin2) --node[anchor=west] {Network} (scope);

%put in background square to highlight the linux SW.
 \begin{pgfonlayer}{background}
    \filldraw [line width=4mm,join=round,red!10]
      (occ.north  -| plugin2.east)  rectangle (plugin2.south  -| plugin2.west);
  \end{pgfonlayer}

\end{tikzpicture}

\caption{Data Flow in RF DAQ: The FPGA produces RF I/Q binary data and sends over PCIe Bus. In the Linux blade, all shown in red, OCC interrupts on PCIe data and stores packets in RAM.  The IOC Driver copies raw packets into a Queue. A  plug-in dequeues the packets and produces floating point I/Q and Magnitude/Phase RF data as EPICS 4 structures, then queues them. Finally a 2nd plug-in accepts the EPICS 4 data and streams over the network.}
\label{fig:flow}
\end{figure}

A plug-in, called the "Parse Plug-in,"  grabs the packet data off the queue and decodes the binary format to create floating point arrays in EPICS4 PVData format. Also, the plug-in does real-time calculations such as converting I and Q into magnitude and phase. The EPICS 4 RF data is then stored into a 2nd queue to be read by a streaming plug-in that serializes and sends the RF data over the network as EPICS4 PVAccess protocol. Clients on the network for real-time display and the File Storage service access the streamed EPICS 4 data over the network.  A diagram of the data flow is shown in Figure \ref{fig:flow}. Additional plug-ins can be added to the software to save RF data in the event of a beam dump.  

When streaming raw ADC data, the 41.5MHz sample rate is too high to send more than one channel at once. Further, because of the slow network connection to the Linux blade (1Gb/sec), the ADC data is sent as blocks of 100k samples of continuous data with missing data between blocks.

\subsection{Real-Time Data Rates}

The AMC streams data via PCIe up to 200MB/sec. There is a
streaming bottle neck in the Linux blade as it has a simple 1Gb Ethernet
connection, and can only stream on the order of 10MB/sec. For saving data to
files, which is a streamed over the network to the on-line data storage service,
the IOC has a queuing system that can temporally store data in case of network
bottle necks. For streaming data to a real time GUI, often data is dropped.

The data rate can be throttled by setting a decimation factor in the FPGA. The DDC
decimates the data by 17, converting 41.5MHz to 2.44MHz. Note that the PID loop
operates at the 2.44MHz sample rate. Because this 2.44MHz sample rate is
typically too high for software streaming, a programmable decimator reduces 
the sample rate by power of 2 decimation factors from 2,4,8..64. Typically we run with
a decimation factor of 8 to produce a streamed sample rate of 305kHz. We can stream I/Q
data over EPICS 4 at this rate for several channels at once. For one channel
the data is reliably continuous. As channels are added, discontinuities are
seen in the real time GUI. Again for saving data to files, the data is
continuous as the DAQ system queues data in memory before saving. For saving data to file,
it is necessary to save short time periods to avoid overrunning the queues in the IOC
software. Typically files with less than 3 seconds of data are saved, leaving files with
continuous data. Because timestamps are saved with data, it is known in data files if
data is not continuous.

%\smartdiagramset{font=\footnotesize,module minimum width=2.2cm,module x sep=2.5,text width=2.0,back arrow disabled=true}
%\smartdiagram[flow diagram:horizontal]{FPGA,PCIe,ADDriver,ParsePlugin,EPICS4}
%\smartdiagram[priority descriptive diagram]{FPGA,PCIe Transfer,DAQ Driver grab packets,DAQ Plugin parse to EPICS4 ,Stream EPICS4 to network,File Store and/or Real-time Display}

\section{GUI and Real-Time Display}

\subsection{LLRF Loop Control}

\begin{figure}
\includegraphics[width=\linewidth]{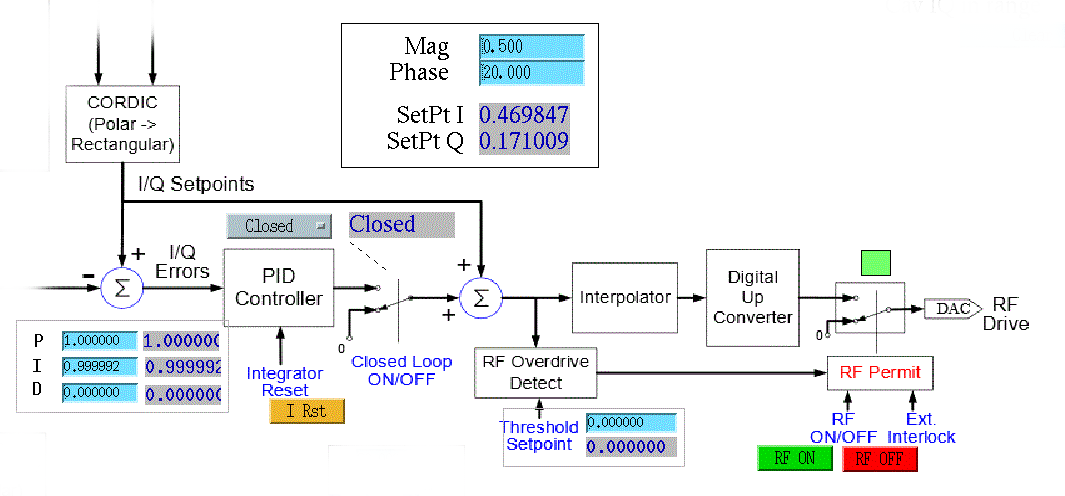}
\caption{Portion of the PID loop user interface.}
\label{fig:llrf}
\end{figure}

The Graphical User Interface (GUI) can be thought of as two components, the
control component and real time display component. The control component is the
interface that allows the set up and control of the digital LLRF hardware. This
interface consists of several user screens with buttons and readouts, and is
displayed by the Motif Editor and Display Manager (MEDM)\cite{medm}. A screen shot of the LLRF feed back loop
screen is shown in Figure \ref{fig:llrf}. 
%in linux to do a gif you convert aaa.gif aaa.png

\subsection{Scope Application}
\begin{figure}
\includegraphics[width=\linewidth]{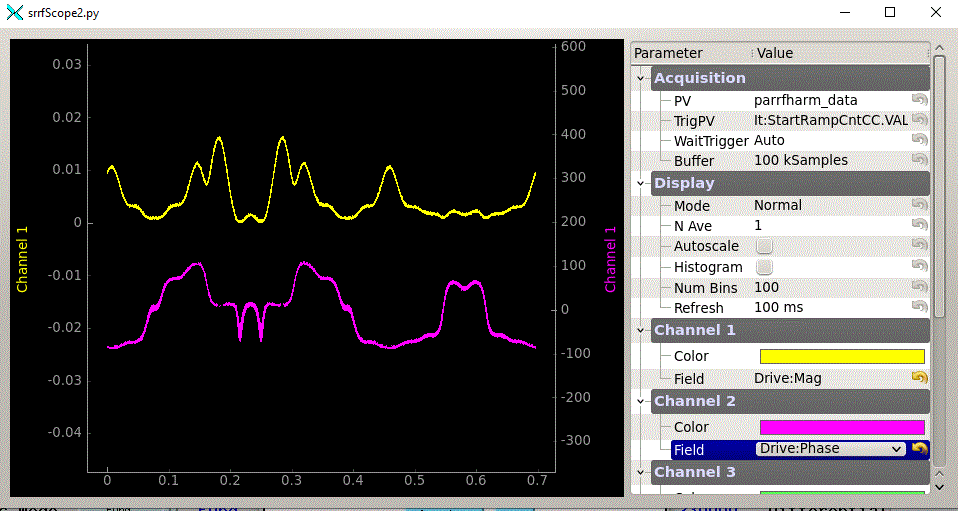}
\caption{Scope Application displaying test waveforms.}
\label{fig:scope}
\end{figure}

The second part of the user interface is the real time display of the RF data.
An application, called the Scope Application, has been written in Python to display
real time waveforms similar to a real hardware oscilloscope. The Scope Application is
an EPICS4 client that receives PVData structures, in the case of LLRF that is
RF I/Q data, and displays the waves on-screen. The Scope Application can up date the
screen at up to 10 frames per sec for real-time performance by using the
PyQtGraph library \cite{pyqtgraph}. Further, the Scope Application can be triggered like
a hardware scope by specifying an EPICS PV. When the EPICS PV changes the scope
is updated. In this way, by using for example a PV that updates for every
Booster cycle, the scope will trigger synchronous with the booster operation.
The jitter of the scope trigger is less than 20ms despite network delays because both the waveform data and trigger PV are timestamped by the APS timing system.
A picture of the Scope Application is shown in Figure \ref{fig:scope}. The Scope is designed
as a software class that can be inherited. In this way, basic functionality
is written into the parent Scope class, and features specific to LLRF are
written into a derived Scope class to make a LLRF-specific Scope. 

The scope has several modes for displaying multiple waveforms in different
colors at once, each with their own scales. Also, data can be displayed 
in the frequency domain, X vs. Y, or as a histogram. Plots can be
averaged in time to create an averaged FFT spectrum.

\section{Data Manipulation}

There would be no point in storing data without some way to manipulate data
and analyze it. At the APS, a structured file format and large set of data
manipulation tools has been developed, called Self Describing Data Set (SDDS) Toolkit\cite{sdds}. SDDS is
a binary file format developed to store accelerator data. The tool set is a
large collection of Linux command line tools for performing data processing,
plotting, and other forms of manipulation. SDDS can also be read by C++,
Python, Matlab, and many other tools for data analysis.

\section{Conclusion and Future Work}

The DAQ software system is currently running at APS and has proven crucial for machine studies and data collection for the APSU. The LLRF software, a small part of the DAQ software, is quite mature and is used for APSU-related data collection and development of the new digital LLRF system. The software will continue to be developed as the APSU progresses.

Future developments include integration of the software with a new time stamping system under development as part of APSU. Also, software specific to various accelerator machines, that is, the Storage Ring, PAR, and Booster, will be developed as APSU progresses. It is hoped that the LLRF software will support faster data rates and high speed networks for more reliable data streaming. More features for storing RF data, especially long term data such as slow storage ring phase drift, are being designed.

\printbibliography

\end{document}